\documentclass[12pt]{article}
\usepackage{graphicx}
\usepackage{epsfig}
\newcommand{\beq}{\begin{equation}}
\newcommand{\eeq}{\end{equation}}
\newcommand{\beqa}{\begin{eqnarray}}
\newcommand{\eeqa}{\end{eqnarray}}
\newcommand{\beqar}{\begin{eqnarray*}}
\newcommand{\eeqar}{\end{eqnarray*}}

\newcommand{\norm}[1]{\raise.3ex\hbox{:}#1\raise.3ex\hbox{:}}

\begin{document}

\setlength{\unitlength}{1mm}

\thispagestyle{empty}

\vspace*{2cm}

\begin{center}
{\bf \LARGE Finite size effects on the Poynting-Robertson effect: a fully general relativistic treatment}\\
\vspace*{1cm}

Jae Sok Oh\footnote{E-mail: ojs@astro.snu.ac.kr}

\vspace*{0.2cm}

{\bf Department of Physics and Astronomy, FPRD,}\\
{\bf Seoul National University, Seoul 151-742, Korea}\\[.5em]

Hongsu Kim\footnote{E-mail: hongsu@Kasi.re.kr}

\vspace*{0.2cm}

{\bf KVN,}\\
{\bf Korea Astronomy and Space Science Institute, Daejeon 305-348, Korea}\\[.5em]

Hyung Mok Lee\footnote{E-mail: hmlee@snu.ac.kr}

\vspace*{0.2cm}

{\bf Department of Physics and Astronomy, FPRD,}\\
{\bf Seoul National University, Seoul 151-742, Korea}\\[.5em]

\vspace{2cm} ABSTRACT
\end{center}
   Even since the first discovery of Poynting and Robertson, the
radiation source has been treated as merely a point. Even in a
very few studies where the size of the source has been taken into
account, the treatment of the problem remained largely
non-relativistic. In the present work, we address the issue of the
finite size effects on the Poynting-Robertson effect in a fully
relativistic manner for the first time.  As a result, the
emergence and the characteristic of the critical point/suspension
orbit can be studied in a systematic and detailed manner.

\vfill \setcounter{page}{0} \setcounter{footnote}{0}
\newpage

\section{Introduction}

   Effects of the radiation field on a test particle orbiting a
radiation source has been studied since the discovery of
Poynting-Robertson effect [1] and [2].
   We begin with the summary of the present status of research in the
literature along this line.
   First, Guess [3] studied the effects of finite size and found that radiation drag
(Poynting-Robertson effect) gets enhanced compared to the case of
point-like radiation source. In his work he confined his
calculation to first order in $v/c$.
   Second, Carrol [4] found that the finite size effect of the radiation source renders
the radiation drag force greater than its counterpart for a
point-like source. His treatment, however, was limited to the
context of special relativity.
   Third, Abramowicz, Ellis, and Lanza [5] constructed
the general relativistic radiation stress-energy tensor describing
the radiation field from the non-rotating radiation source with
finite radius size for the first time and solved the equation of
motion to discover the existence and the location of a critical
point at which all segments of accretion flow are captured. As
their concern was to describe jets and outflows, they limited the
particle's motion to one-dimensional radial motion alone.
   Fourth, Miller and Lamb [6] and [7] considered and
solved the azimuthal as well as radial component of the equation
of motion. They, however, failed to notice the emergence of and
study the nature of critical point as they considered the set-up
in which the luminosity of the radiation source is well below the
Eddington critical value.
   Interestingly enough, lastly, Bini, Jantzen, and
Stella [8] studied general relativistic version of the
Poynting-Robertson effect by considering and solving all
3-components (time, radial, and azimuthal) of the equation of
motion.  Their set-up, however, is not identical to that in our
present work in that they employed the radiation stress-energy
tensor describing the radiation field from a point source.  They
ignored the finite size of the radiation source.

   In the present work, in order to address general
relativistic version of finite size effects on the
Poynting-Robertson effect in a rigorous and complete manner, we
employed the radiation stress-energy tensor first constructed by
Abramowicz, Ellis, and Lanza [5] to explore the detailed impact of
the radiation forces on the test particle's trajectories.

   In section 2 we describe the derivation of the equation of motion
of the test particle in the Schwarzschild space-time background
outside a non-rotating, isotropic radiating spherical star with
finite size. In section 3, we present the results of the numerical
integrations along with the our analytical results. Finally in
section 4, we summarize and discuss our results.

\section{Equations of motion}

   To derive the equations of motion that govern the
orbital motion of the test particle in the radiation field from
the spherical star emitting the radiation isotropically, we start
with Schwarzschild spacetime,
\begin{eqnarray}
   ds^2 &=& g_{\mu\nu}dx^{\mu}dx^{\nu} \nonumber \\
        &=& -(1- 2M/r)dt^{2} \,+\, (1- 2M/r)^{-1}dr^{2} \,+\, r^2(d\theta^{2}
        \,+\, {\sin}^{2} \theta d \phi^{2}).
\end{eqnarray}
    We work in the geometric units, where $G = 1 = c$ ($G$
is the gravitational constant and $c$ is the speed of light).
    The equations of motion is given by
\begin{eqnarray}
   a^{\alpha} = \frac {f^{\alpha}}{m}
\end{eqnarray}
    where $f^{\alpha}$ denotes a non-gravitational force (for example,
radiation force) exerted by the radiation (or luminosity) on the
test particle, $m$ is the rest mass of the test particle, and
\begin{eqnarray}
  a^{\alpha} = \frac {dU^{\alpha}}{d\tau} \; + \;
  \Gamma^{\alpha}_{\mu\nu} U^{\mu}U^{\nu}
\end{eqnarray}
is the acceleration, where $U^{\alpha}$ is the four-velocity of
the particle, $\Gamma^{\alpha}_{\mu\nu} = \frac{1}{2}
g^{\alpha\beta} (g_{\beta\mu ,\nu} + g_{\beta\nu ,\mu} - g_{\mu\nu
,\beta})$ is the Affine connection, and comma (,) denotes partial
derivatives.

    If we assume that the cross-section $\sigma$ of the test particle is independent of
energy (frequency) and direction of the photon, the
non-gravitational force $f^{\alpha}$ due to scattering of photon
from the star is given by [9],
\begin{eqnarray}
   f^{\alpha} = \sigma F^{\alpha},
\end{eqnarray}
    where $F^{\alpha}$ shown below is the quantity transformed from the radiation
energy flux $T_{co}^{\hat i\hat 0}$ measured in the particle's
rest frame (which is comoving with the particle) using the
orthonormal tetrad $\tilde {e}^{\alpha}_{\hat i}$ associated with
the particle's rest frame as follows,
\begin{eqnarray}
   F^{\alpha} &=& \tilde {e}^{\alpha}_{\hat i} T_{co}^{\hat i\hat
   0} \nonumber \\
              &=& - h^{\alpha}_{\beta} T^{\beta\sigma} U_{\sigma},
\end{eqnarray}
    where $h^{\alpha}_{\beta} = \delta^{\alpha}_{\beta} +
U^{\alpha}U_{\beta}$ is the projection tensor, and
$T^{\beta\sigma}$ is the radiation stress-energy tensor in the lab
frame.

    According to [5], the components of the radiation stress tensor
$T^{\hat\alpha \hat\beta}$ measured in LNRF (Locally Non-Rotating
Frame; [10, 11]) can be written as,
\begin{eqnarray}
   T^{\hat t\hat t} &=& 2\pi I(r)(1-\cos\alpha) \\
   T^{\hat t\hat r} &=& \pi I(r)\sin^{2}\alpha \\
   T^{\hat r\hat r} &=& \frac{2}{3}\pi I(r)(1-\cos^{3}\alpha)\\
   T^{\hat\theta \hat\theta} &=& \frac{\pi}{3}
                    I(r)(\cos^{3}\alpha - 3\cos\alpha + 2) \\
   T^{\hat\phi \hat\phi} &=& \frac{\pi}{3}
                    I(r)(\cos^{3}\alpha - 3\cos\alpha + 2),
\end{eqnarray}
    where $\alpha$ is an apparent viewing angle of the star seen by an
observer (FIDO; fiducial observer) at rest in the LNRF and is
given by $\sin \alpha = \left(\frac{R}{r} \right) {\left( \frac{1
- 2M/r}{1 - 2M/R}\right)}^{1/2}$ (see [5]) for the radius of the
star $R \geq 3M$, and $I(r)$ is the frequency-integrated specific
intensity at the radial position $r$ and can be expressed as (see
Appendix A in the [7])
\begin{eqnarray}
   I(r) = \frac{(1-2M/R)}{(1-2M/r)^2} \frac{m M}{\pi\sigma R^2}
          \left(\frac{L^{\infty}}{L_{Edd}^{\infty}}\right),
\end{eqnarray}
where $L^{\infty}$ is the luminosity at infinity and
$L^{\infty}_{Edd}$ is the Eddington critical luminosity, which is
given by $L^{\infty}_{Edd} = 4{\pi}mM/\sigma$.

    By transforming $T^{\hat\mu \hat\nu}$
using the tetrad $e^{\alpha}_{\hat\mu}$ associated with the LNRF,
we can obtain the components of the radiation stress tensor
$T^{\alpha\beta}$,
\begin{eqnarray}
   T^{\alpha\beta} = e^{\alpha}_{\hat\mu} e^{\beta}_{\hat\nu}
   T^{\hat\mu\hat\nu},
\end{eqnarray}
    where the tetrad associated with LNRF in the
Schwarzschild coordinates is given by,
\begin{eqnarray}
   e^{\hat 0} &=& (1 - 2M/r)^{1/2}dt, \nonumber \\
   e^{\hat 1} &=& (1 - 2M/r)^{-1/2}dr, \\
   e^{\hat 2} &=& rd \theta, \nonumber \\
   e^{\hat 3} &=& r \sin \theta d \phi. \nonumber
\end{eqnarray}

Finally, the radiation stress-energy tensor describing the
radiation field from the radiation source of finite size ($R \geq
3M$) can be rewritten as,
\begin{eqnarray}
   T^{tt} &=& \left(\frac{m}{\sigma}\right)
              \frac{2}{(1+\cos\alpha)}
              \left(\frac{M}{r^2}\right)
              \left( 1-\frac{2M}{r} \right)^{-2}
              \left(\frac{L^{\infty}}{L_{Edd}^{\infty}}\right) \\
   T^{tr} &=& \left(\frac{m}{\sigma}\right)
              \left(\frac{M}{r^2}\right)
              \left( 1-\frac{2M}{r} \right)^{-1}
              \left(\frac{L^{\infty}}{L_{Edd}^{\infty}}\right) \\
   T^{rr} &=& \left(\frac{m}{\sigma}\right)
              \frac{2(1+\cos\alpha+\cos^{2}\alpha)}{3(1+\cos\alpha)}
              \left(\frac{M}{r^2}\right)
              \left(\frac{L^{\infty}}{L_{Edd}^{\infty}}\right) \\
   T^{\theta\theta} &=& \left(\frac{m}{\sigma}\right)
              \frac{(2-\cos\alpha-\cos^{2}\alpha)}{3(1+\cos\alpha)}
              \left(\frac{M}{r^4}\right)
              \left( 1-\frac{2M}{r} \right)^{-1}
              \left(\frac{L^{\infty}}{L_{Edd}^{\infty}}\right) \\
   T^{\phi\phi} &=& \left(\frac{m}{\sigma}\right)
              \frac{(2-\cos\alpha-\cos^{2}\alpha)}{3(1+\cos\alpha)}
              \left(\frac{M}{r^4 \sin^{2}\theta}\right)
              \left( 1-\frac{2M}{r} \right)^{-1}
              \left(\frac{L^{\infty}}{L_{Edd}^{\infty}}\right).
\end{eqnarray}

When a radiation source is point-like, the apparent viewing angle
($\alpha$) measured by FIDO is equal to zero, thus $\cos\alpha =
1$. Inserting this $\cos\alpha = 1$ into above equations (14)
through (18) gives the radiation stress-energy tensor for
point-like radiation source, which is exactly identical to the
radiation stress-energy tensor employed by Bini, Jantzen, and
Stella [8].

    When the radius ($R$) of the radiation source is smaller than the
photon sphere ($3M$), i,e., $R \leq 3M$, the apparent viewing
angle is given by,
\begin{eqnarray}
     \sin\alpha = 3\sqrt{3} \left(\frac{M}{r}\right)
     \left(1-\frac{2M}{r}\right)^{1/2}.
\end{eqnarray}
Therefore, the radiation stress-energy tensor describing the
radiation field from the radiation source which has a radius
smaller than the photon sphere have no dependence of the source's
radius $R$. Thus, such radiation sources can be treated as a
radiation source with $R=3M$(the radius of the photon sphere).

    We can restrict the orbital motion of the particle to the equatorial plane
($\theta = \frac{\pi}{2}, U_{\theta} = 0$), without loss of
generality owing to the spherical symmetry of non-rotating
spherical star, thus the equation of motion finally can be
decomposed into each components as follows;

\begin{eqnarray}
   \frac{dU_t}{d\tau} &=& \left(\frac{M}{r^2}\right)
                          \left[1-2\left(1-\frac{2M}{r}\right)^{-1}U^{2}_{t}\right]
                          \left(\frac{L^{\infty}}{L^{\infty}_{Edd}}\right)
                          U_{r}
                           \\
                      &-& \frac{(8+2\cos\alpha+2\cos^{2}\alpha)}{3(1+\cos\alpha)}
                          \left(\frac{M}{r^2}\right)
                          \left(\frac {L^{\infty}}{L^{\infty}_{Edd}}\right)
                          U_{t}U^{2}_{r}
                          \nonumber \\
                      &-& \frac{(8-\cos\alpha-\cos^{2}\alpha)}{3(1+\cos\alpha)}
                          \left(\frac{M}{r^4}\right)
                          \left(1-\frac{2M}{r}\right)^{-1}
                          \left(\frac {L^{\infty}}{L^{\infty}_{Edd}}\right)
                          U_{t}U^{2}_{\phi}
                          \nonumber
\end{eqnarray}

\begin{eqnarray}
   \frac{dU_r}{d\tau} &=& - \left(\frac{M}{r^2}\right)
                            \left(1-\frac{2M}{r}\right)^{-1}
                          - \left(\frac{2M}{r^2}\right)U^{2}_{r}
                          + \left(\frac{1}{r^3}\right)
                            \left(1-\frac{2M}{r}\right)^{-1}
                            \left(1-\frac{3M}{r}\right)U^{2}_{\phi}
                            \\
                      &-&   \left(\frac{M}{r^2}\right)
                            \left(1-\frac{2M}{r}\right)^{-2}
                            \left[1+2\left(1-\frac{2M}{r}\right)U^{2}_{r}\right]
                            \left(\frac{L^{\infty}}{L^{\infty}_{Edd}}\right)
                            U_{t}
                            \nonumber \\
                      &-& \frac{(8+2\cos\alpha+2\cos^{2}\alpha)}{3(1+\cos\alpha)}
                          \left(\frac{M}{r^2}\right)
                          \left(1-\frac{2M}{r}\right)^{-1}
                          \left[1+\left(1-\frac{2M}{r}\right)U^{2}_{r}\right]
                          \left(\frac {L^{\infty}}{L^{\infty}_{Edd}}\right)
                          U_{r}
                          \nonumber \\
                      &-& \frac{(8-\cos\alpha-\cos^{2}\alpha)}{3(1+\cos\alpha)}
                          \left(\frac{M}{r^4}\right)
                          \left(1-\frac{2M}{r}\right)^{-1}
                          \left(\frac {L^{\infty}}{L^{\infty}_{Edd}}\right)
                          U_{r}U^{2}_{\phi}
                          \nonumber
\end{eqnarray}

\begin{eqnarray}
   \frac{dU_{\phi}}{d\tau} &=& -\left(\frac{2M}{r^2}\right)
                            \left(1-\frac{2M}{r}\right)^{-1}
                            \left(\frac{L^{\infty}}{L^{\infty}_{Edd}}\right)
                            U_{t}U_{r}U_{\phi}
                            \\
                       &-&  \frac{(8-\cos\alpha-\cos^{2}\alpha)}{3(1+\cos\alpha)}
                            \left(\frac{M}{r^2}\right)
                            \left(1-\frac{2M}{r}\right)^{-2}
                            \left(\frac {L^{\infty}}{L^{\infty}_{Edd}}\right)
                            U^{2}_{t}U_{\phi}
                            \nonumber \\
                       &-&  \cos\alpha
                            \left(\frac{M}{r^2}\right)
                            \left(\frac {L^{\infty}}{L^{\infty}_{Edd}}\right)
                            U^{2}_{r}U_{\phi}.
                            \nonumber
\end{eqnarray}

  We are ready to envisage the features of the solution to the
equation of motion.
  First, equation (21), radial component of
equation of motion, includes four lines: the three terms in the
first line is not relevant to the radiation. as overall sign of
the second line is positive definite, this line can be identified
with the outward radiation pressure gradient and it should be
noted that the second line has no dependence of the finite size of
the radiation source. Both third line and fourth line can be
regarded as being responsible for the radial radiation drag force
as their overall sign are negative definite. As the third line
depends on the radial velocity component alone, if we take into
account only radial motion, the radial radiation drag is
attributable to the third line. As we can know from the finite
size dependent part
($\frac{(8+2\cos\alpha+2\cos^{2}\alpha)}{3(1+\cos\alpha)}$) of the
third line, the radial radiation drag at the surface of the
radiation source ($\cos\alpha = 0$) gets 4/3 times larger than
that of the point-like source ($\cos\alpha = 1$), which is
identical with the results in the study of Guess [3] and Carrol
[4]. However, their study failed to mention about the our fourth
line radiation drag which depends on both the radial velocity and
the azimuthal velocity. The radial radiation drag in the fourth
line gets 8/3 time large compared to the point-like source.
Second, equation (22), azimuthal component of equation of motion,
is composed of three lines: the first line depends on both the
radial and the azimuthal velocity and its overall sign relies on
the direction of the radial motion of the particle. When the
particle is in inward motion, as its overall sign is negative
definite, the first line can be regarded as the azimuthal
radiation drag, but if the motion of the particle is changed to
outward one, as its overall sign is changed to positive definite,
it can be identified with the radiation counter-drag. As the
second line has a dependence of the azimuthal velocity only and
its overall sign is negative definite, this second line can be
regarded as the azimuthal radiation drag exerted on the particle
in circular motion.  The finite size dependent part in the second
line indicates that the azimuthal radiation drag at the surface of
the radiation source gets 8/3 times larger than that of the
point-like source, which also is identical with the results in the
study of Guess [3] and Carrol [4]. Finally, the third line depends
on both the radial and the azimuthal velocity and its overall sign
is negative definite, thus this third line can be also identified
with the radiation drag. Interestingly, the finite size of the
radiation source leads to smaller radiation drag than a point-like
source.


\section{Results of numerical integration}

   For the sake of convenience, we introduce luminosity parameter $L
\equiv \frac{L^{\infty}}{L_{Edd}^{\infty}}$.

\begin{figure}
\begin{center}
\epsfysize=8truecm \epsffile{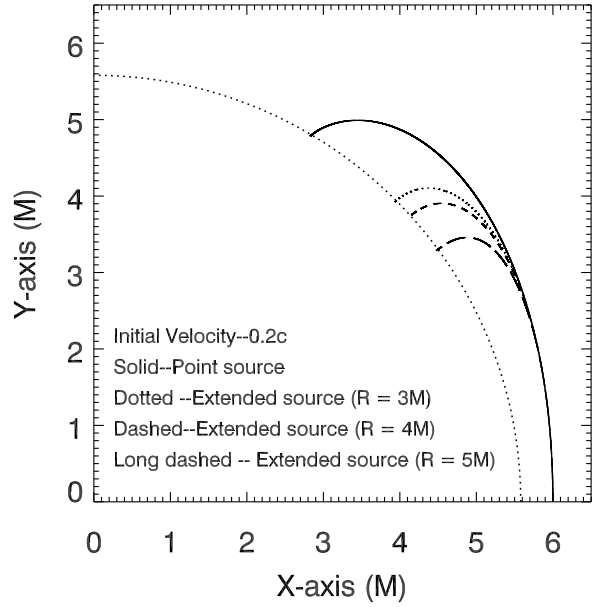}
\end{center}
\caption{shows the ways the test particle arrives and enters onto
the suspension orbit. 4-trajectories provided exhibit as the
radiation source gets larger and hence it's surface gets wider,
the Poynting-Robertson effect gets more and more manifest reducing
the test particle's azimuthal speed more and more remarkably. The
luminosity parameter of the radiation sources are the same as
$L=0.8$ and dotted circle denotes ``suspension orbit".}
\end{figure}

\begin{figure}
\begin{center}
\epsfysize=8truecm \epsffile{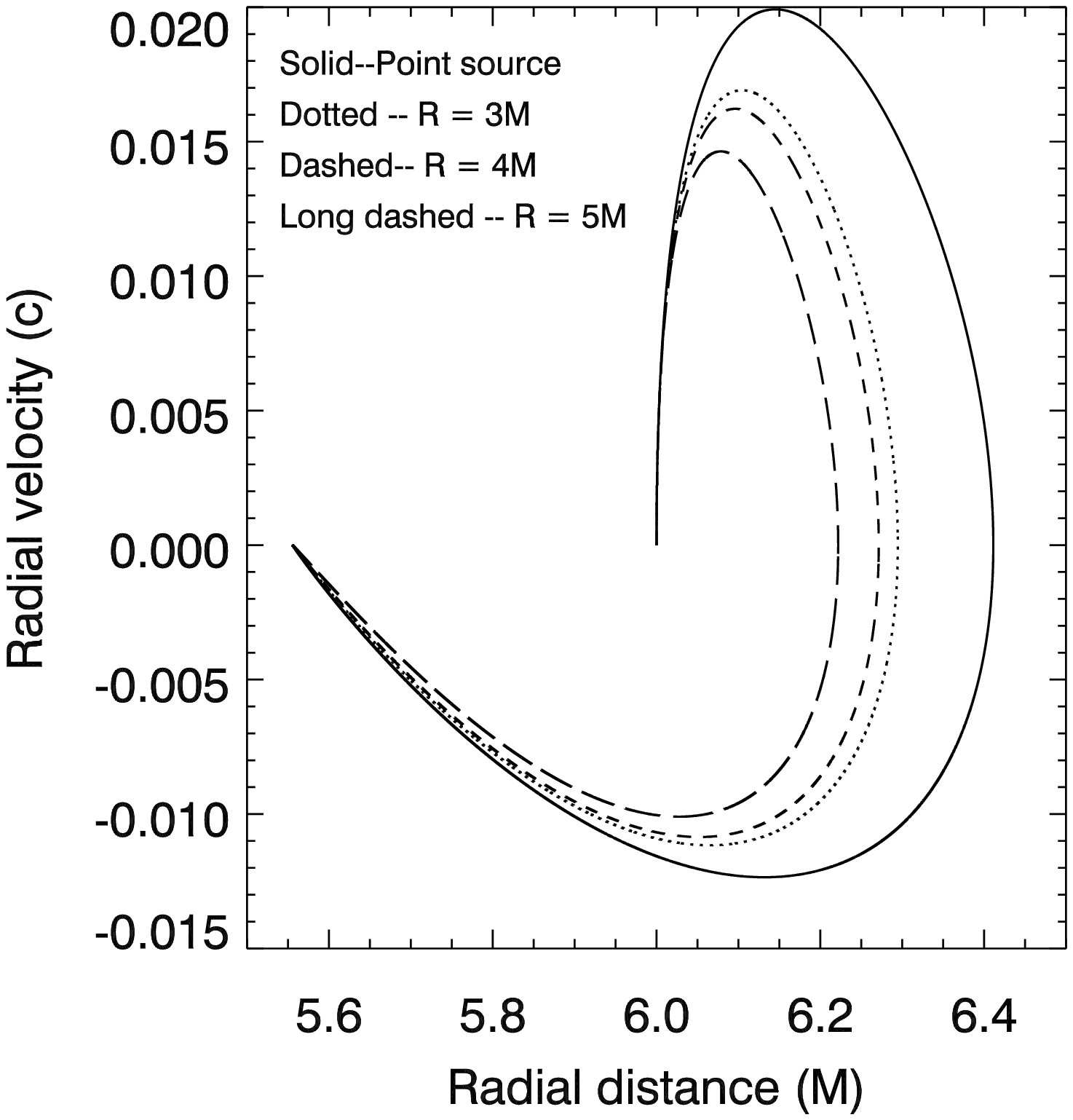}
\end{center}
\caption{demonstrates the radial velocity profiles of the
4-trajectories of the test particle given in Figure 1.}
\end{figure}

\begin{figure}
\begin{center}
\epsfysize=8truecm \epsffile{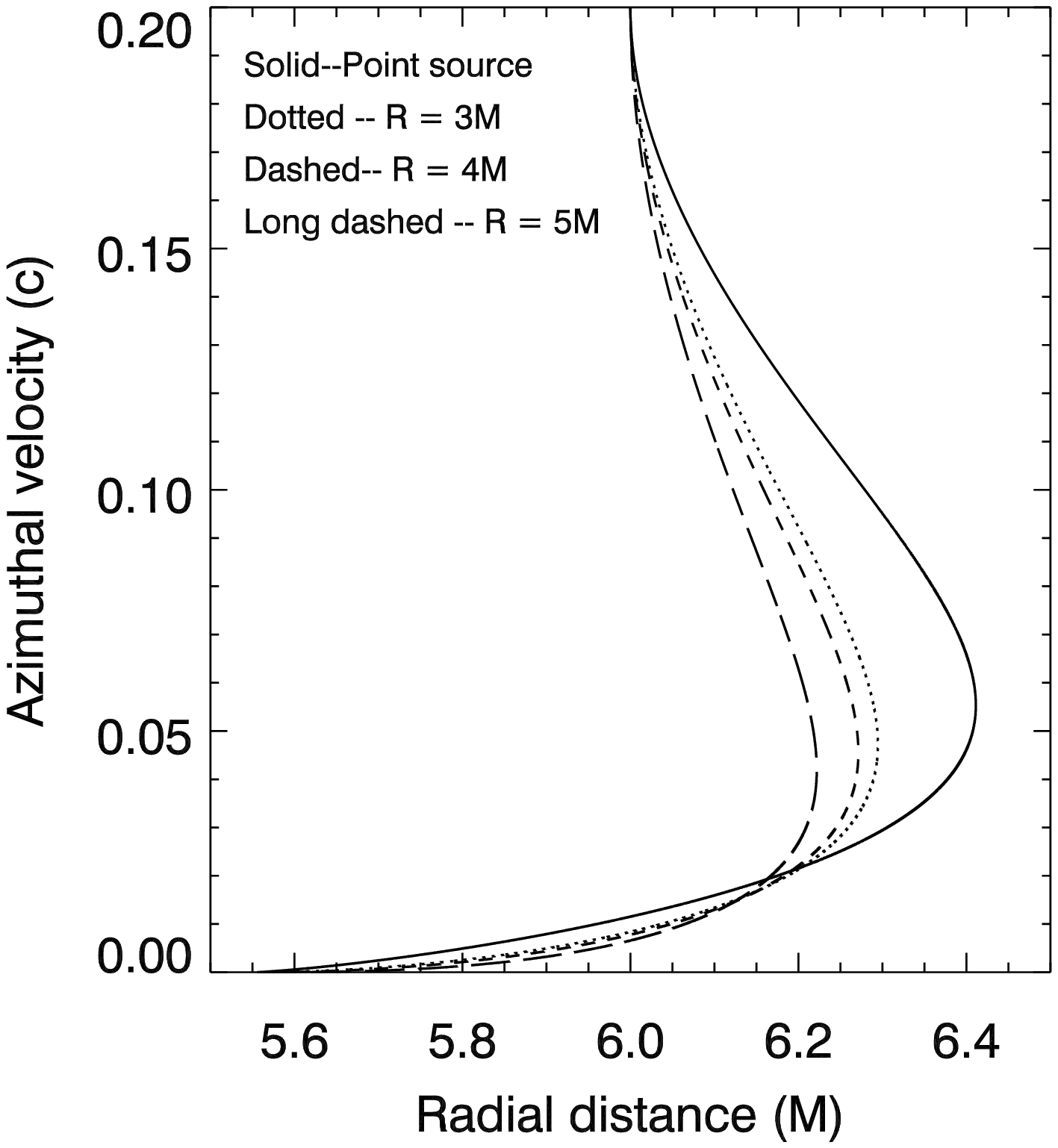}
\end{center}
\caption{demonstrates the azimuthal velocity profiles of the
4-trajectories of the test particle given in Figure 1.}
\end{figure}

   Figure 1 shows the ways the test particle arrives and enters
onto the dotted circle. The test particle arriving at the dotted
circle seems to be suspended and not to move any more there, thus
hereinafter we will refer to the dotted circle as ``suspension
orbit".  The starting points of the particle are the same as
$(6M,0)$ in Cartesian coordinates (X,Y). Initial azimuthal
velocities are fixed to be 0.2c and its direction is
counterclockwise. Also, the luminosity of the radiation sources
are the same as $L=0.8$. Radial distance of the ``suspension
orbit" is roughly $r_{so} \simeq 5.56M$. Solid line denotes the
trajectory of the test particle when the radiation source is
point-like, and dotted, dashed, and long dashed lines denote the
trajectories of the particle when the radiation sources have the
radii of 3M, 4M, and 5M, respectively. As it is mentioned above
section II, it should be noted that the radiation source with
radius smaller than $3M$ (photon sphere) can be treated as one
with radius of $3M$.

   As shown in Figure 1, as the size (radius) of the radiation source
gets larger, the ``suspension orbit" stays the same (location) but
the test particle arrives at the ``suspension orbit" more and more
quickly.  Unlike the case of the point source where the test
particle hits the ``suspension orbit" at some finite angle, when
the source has finite size, the test particle always strikes the
``suspension orbit" nearly perpendicularly.

   Figures 2 and 3 demonstrate the radial and azimuthal velocity
profiles of the 4-trajectories of the test particle given in
Figure 1, respectively.  As shown in Figures 2 and 3, around the
``suspension orbit", as the size of the radiation source gets
larger, the radial and azimuthal velocities are more and more
quickly reduced and finally vanish at the ``suspension orbit".
Thus, from Figures 1 through 3 we can find out that the radial and
azimuthal velocities of the test particle at the ``suspension
orbit" are, respectively, equal to zero and the rate of change of
the radial and azimuthal velocities also vanish, respectively,
hence the test particle comes to a complete rest and does not move
any more there. In other word, the test particle there satisfies
the following conditions,

\begin{eqnarray}
        U_{r} &=& U_{\phi} = 0 \nonumber \\
    \frac{dU_{t}}{d\tau} &=& \frac{dU_{r}}{d\tau} =
    \frac{dU_{\phi}}{d\tau} =0. \nonumber
\end{eqnarray}

   For $U_{r} = U_{\phi} = 0$, the normalization condition
$g^{\mu\nu}U_{\mu}U_{\nu} = -1$ yields $U_{t} = -
(1-{2M}/r)^{1/2}$. Equations (20) and (22) satisfy the above
conditions, and Plugging this $U_{t}= - (1-{2M}/r)^{1/2}$ together
with $\frac{dU_{r}}{d\tau} = 0$ into equation (21) gives,

\begin{eqnarray}
  0 = - \left(\frac{M}{r^2}\right) \left(1-\frac{2M}{r}\right)^{-1/2}
      + \left(\frac{M}{r^2}\right) \left(1-\frac{2M}{r}\right)^{-1}
        \left(\frac{L^{\infty}}{L^{\infty}_{Edd}}\right).
\end{eqnarray}

   The solution of Equation (23) gives the radial distance $r_{so}$ of the
``suspension orbit",

\begin{eqnarray}
  r_{so} = {2M}\left[{1-\left(\frac{L^{\infty}}{L^{\infty}_{Edd}}\right)^2}\right]^{-1}
\end{eqnarray}
which is equivalent to the radial position of the critical point
in Abramowicz et al.[5] and Bini et al.[8], and is determined by
$\left(\frac{L^{\infty}}{L^{\infty}_{Edd}}\right)$ alone.

  Since first term on the right-handed side in equation (23) denotes
gravitational inward force exerted on the test particle at
complete rest in the ``suspension orbit" and second term refers to
the outward radiation pressure there, equation (23) means that the
gravitational inward force at the ``suspension orbit" exactly
balance the outward radiation pressure there.

  It is interesting that among the entries/components of the
radiation stress-energy tensor, the (t,t)-component, which is the
radiation energy density and the (r,r)-component, which is the
radiation pressure, exerts only on the moving particle, while the
only (t,r)-component, which is the pressure gradient (force)
exerts on both the stationary particle and the moving particle. As
we already mentioned above, the (t,t) and (r,r) components of our
choice of the stress-energy tensor are different from those of
Bini et al.[8] whereas the (t,r) component of ours is exactly same
as that of Bini et al.[8].

  As a result, although the trajectory of the test particle before
it reaches the ``suspension orbit" in our study is manifestly
different from that in the study of Bini et al.[8], once the test
particle arrives at the ``suspension orbit", the outward radiation
pressure gradient that balance the inward gravitational force in
our study is the same as that in the study of Bini et al.[8] and
this is precisely why our radial distance of the ``suspension
orbit'' and theirs turn out to be the same.

\section{Concluding remarks}

   Although since the remarkable discovery first made by Poynting
and Robertson, a number of study on this Poynting-Robertson effect
have been performed, in almost all the works the luminous
radiation source has been treated as being point-like. This could
be due to their particular aims of investigations in which they
just tried to eliminate unnecessary complexity which has little to
do with essence of their studies.  Largely, however, such
simplification or reduction of the set-up can be attributed to the
technical and computational difficulties.  In order to render the
set-up for the careful and detailed investigation of the
Poynting-Robertson effect to be more realistic and practical,
obviously one has to take into account the finite size of the
radiating source.  Even in a very few works in the literature
addressing this issue of finite size, the treatment of the issue
has been very limited. For instance, Abramowicz et al. [5]
attempted general relativistic and analytical approach, but they
considered radial equation of motion alone.  Next, Miller and Lamb
[6,7] performed general relativistic and numerical approach, but
they considered angular equation of motion alone.  To summarize,
even these few works addressing the issue of finite size effect
for the Poynting-Robertson effect have remained essentially
incomplete.  Along this line, our present work can be regarded as
comprehensive treatment of the finite size effects on the
Poynting-Robertson effect in a fully general relativistic context
for the first time ever.   As a result of this comprehensive
treatment of the problem, we were successful in realizing the
existence of critical point/suspension orbit and in understanding
the detailed characteristics of this suspension orbit.   It is our
hope, therefore, that the present study of ours could solve as a
ground work on which further serious studies addressing the issue
of finite size effects on the Poynting-Robertson effect can be
successfully carried out.

\vspace*{1cm}

\section*{Acknowledgments}

   J.S.O. acknowledges the support of the BK21 program to SNU.

\newpage

\begin{center}
{\rm\bf References}
\end{center}

[1] Poynting J. H., Phil. Trans. R. Soc. {\bf 203}, 525 (1903).\\

[2] Robertson H. P., M.N.R.A.S {\bf 97}, 423 (1937).\\

[3] Guess, A. W., Astrophys. J. {\bf 135}, 855 (1962).\\

[4] Carrol, D. L., Astrophys. J. {\bf 348}, 588 (1990).\\

[5] Marek A. Abramowicz, George F. R. Ellis, and Antonio Lanza, Astrophys. J. {\bf 361}, 470 (1990).\\

[6] M. Coleman Miller and Frederick K. Lamb, Astrophys. J. {\bf 413}, L43 (1993).\\

[7] M. Coleman Miller and Frederick K. Lamb, Astrophys. J. {\bf 470}, 1033 (1996).\\

[8] Donato Bini, Robert T Jantzen, and Luigi Stella, Class. Quantum Grav. {\bf 26}, 055009 (2009).\\

[9] Frederick K. Lamb and M. Coleman Miller, Astrophys. J. {\bf 439}, 828 (1995).\\

[10] J. M. Bardeen, Astrophys. J. {\bf 162}, 71 (1970).\\

[11] J. M. Bardeen, W. H. Press, and S. A. Teukolsky, Astrophys. J. {\bf 178}, 347 (1972).\\

\end{document}